\documentclass[tablecaption=bottom, wcp]{jmlr} %

\usepackage[USenglish]{babel}
\usepackage[compact]{titlesec}
\usepackage[symbol]{footmisc}

\usepackage[acronym,smallcaps,nowarn,section,nogroupskip,nonumberlist]{glossaries}
\glsdisablehyper

\usepackage{csquotes}

\makeatletter
\def\set@curr@file#1{\def\@curr@file{#1}} %
\makeatother
\usepackage[load-configurations=version-1]{siunitx} 

\usepackage{IEEEtrantools}
\usepackage{color, xcolor}

\usepackage{bbm}

\renewcommand{\d}{\mathrm{d}}

\newacronym{abc}{abc}{approximate Bayesian computation}
\newacronym{mmd}{mmd}{maximum mean discrepancy}
\newacronym{pmcmc}{pmcmc}{particle Markov chain Monte Carlo}
\newacronym{mcmc}{mcmc}{Markov chain Monte Carlo}
\newacronym{ess}{ess}{effective sample size}
\newacronym{gbilfi}{gb-lfi}{general Bayes for likelihood-free inference}
\newacronym{smc}{smc}{sequential Monte Carlo}
\newacronym{gbi}{gbi}{generalized Bayesian inference}

 \usepackage{hyperref}
 \hypersetup{
     colorlinks = true,
     linkcolor = black,
     anchorcolor = blue,
     citecolor = blue,
     filecolor = blue,
     urlcolor = blue
     }
 
\usepackage[load-configurations=version-1]{siunitx} %

\jmlrproceedings{AABI 2020}{3rd Symposium on Advances in Approximate Bayesian Inference, 2020}

\title[Generalized posteriors in abc]{Generalized Posteriors in Approximate Bayesian Computation}

\author{
\Name{Sebastian M Schmon\nametag{\thanks{equal contribution}}}  \Email{sebastianschmon@improbable.io} \\
\addr Improbable \\
\Name{Patrick W Cannon\nametag{\footnotemark}}  
\Email{patrickcannon@improbable.io}\\
\addr Improbable \\
\Name{Jeremias Knoblauch} \Email{j.knoblauch@warwick.ac.uk} \\
\addr University of Warwick \& The Alan Turing Institute
}

\usepackage{mathtools}
\newcommand{\defeq}{\vcentcolon=}

\begin{document}

\maketitle

\begin{abstract}
Complex simulators have become a ubiquitous tool in many scientific disciplines, providing high-fidelity, implicit probabilistic models of natural and social phenomena. 
Unfortunately, they typically lack the tractability required for conventional statistical analysis.
\Gls{abc} has emerged as a key method in \textit{simulation-based inference},  wherein the true model likelihood and  posterior are approximated using samples from the simulator. 
In this paper, we draw connections between \gls{abc} %
and \gls{gbi}. %
First, we re-interpret the accept/reject step in \gls{abc} as an implicitly defined error model.
We then argue that these implicit error models will invariably be misspecified. 
While \gls{abc} posteriors are often treated as a necessary evil for approximating the standard Bayesian posterior, this allows us to re-interpret \gls{abc} as a potential robustification strategy.
This leads us to suggest the use of \gls{gbi} within \gls{abc}, a use case we explore empirically.

\end{abstract}

\section{Introduction}
\label{sec:intro}

\glsresetall

\Gls{abc} is a family of techniques for computing approximate Bayesian posteriors in the presence of intractable likelihoods \citep[see e.g. ][]{beaumont2002approximate, fan2018abc, beaumont2019approximate}.
Accordingly, such methods are particularly attractive when one is confronted with complex physical \textit{simulators}.
Suppose we are given a probabilistic simulator $f(\cdot \mid \theta)$ parameterized by $\theta \in \Theta$ and defining a probability distribution on $\mathcal{X}$.
Assuming that $\int_{\mathcal{X}}f(x\mid \theta)\, \d x = 1$, this simulator also defines a likelihood.
If we have observations $y$ and a prior belief $p(\theta)$, the standard Bayesian approach is to update our belief according to Bayes' rule,
\begin{IEEEeqnarray}{rCl}
    p(\theta \mid y) & \propto & f(y \mid \theta)p(\theta).
    \nonumber
\end{IEEEeqnarray}
Clearly, this approach is infeasible for general simulation engines. While we can sample $x \sim f(\cdot \mid \theta)$ for any $\theta \in \Theta$, evaluating $f(y \mid \theta)$ for a given $y$ and $\theta$ is in general not possible. 
The picture is further complicated by misspecification: the simulator $f(\cdot \mid \theta)$ will seldom  describe the true data generating process $y \sim p^*(\cdot)$ perfectly. %
While many approaches exist for tackling this \textit{intractable likelihood} problem \citep[see e.g.][for a recent overview]{cranmer2020frontier}, almost all of them rely on the assumption that the likelihood model defined through the simulator is correctly specified.
In the following, we will show how \gls{gbi} can be used to design \gls{abc} algorithms robust to misspecification.

Operationally, \gls{abc} describes the following algorithm: first, sample $\theta \sim p(\theta)$; next, sample $x\mid \theta \sim f(x\mid\theta)$ by running the simulator with parameter $\theta$; lastly, weight the disagreement between $x$ and the observed data $y$ by comparing low-dimensional summary statistics, $\eta(x)$ and $\eta(y)$, as judged by some probability kernel function $K_h(\|\eta(x)-\eta(y)\|)$ with bandwidth $h$.
This induces an augmented distribution $p(\theta, x \mid y)$, a marginal of which yields an approximation to the Bayesian posterior
\begin{IEEEeqnarray}{rCl}
    p(\theta \mid y) 
    \approx p_{\gls{abc}}(\theta \mid y) 
    \defeq \int p(\theta, x \mid y) \, \d x
    \propto \int K_h(\|\eta(x)-\eta(y)\|)f(x|\theta)p(\theta) \, \d x.
    \label{eq:ABC_generic}
\end{IEEEeqnarray}
The logic in \eqref{eq:ABC_generic} is immediate: the smaller the bandwidth $h$ of the kernel, the more accurate our approximation to the true posterior will be.
On the other hand, smaller bandwidths  result in down-weighting or discarding more of the simulator outputs.
Further, one needs to choose a summary statistic $\eta$ that is ``as sufficient as possible'' to optimally use all information in the data. 
This tension between the theoretically optimal and the practically feasible is a recurrent theme within the literature on \gls{abc} \citep[e.g.][]{Blum_2013}.

In the current paper, we use recent insights from %
\gls{gbi} \citep[see e.g.][]{safeBayes, bissiri2016general, JackDivergences, PACBayes, GVI} to form a new perspective on these issues.
The first key step is the reinterpretation of the kernel $K_h$ as the \textit{measurement error model} for the true data $y$.
Unlike the standard interpretation of $K_h$ in \gls{abc}, this suggests that larger bandwidths $h$ could actually be \textit{beneficial} for more robust inferences. 
More precisely, when the simulator is a poor description of $y$ for any value of $\theta$, an explicit model of the noise may robustify inference.
As the noise models themselves will also be misspecified, we further draw a natural connection to \gls{gbi}: using generalizations of the Bayesian principle, one can circumvent the drawbacks of having a substantially misspecified noise model.
In summary, rather than treating \gls{abc} as a necessary evil for approximating the standard Bayesian posterior, we re-interpret it as a flexible robustification strategy.

\section{Bayesian calibration of stochastic simulators}
Traditional approaches for calibrating stochastic simulators assume that the model is well-specified, inherent in the commonly stated aim to approximate the posterior of the simulator 
$$
    p(\theta \mid y) \propto f(y \mid \theta)p(\theta).
$$
However, if $y$ is a real-world observation, it is rarely safe to assume it is drawn from the simulator. 
On the contrary, it is reasonable to assume that the simulator is misspecified to a significant degree---either because of measurement error in the data collection process or because of the modelling challenge of capturing the true dynamics. 
The latter is of particular importance in the simulation of e.g. social dynamics and economic models, where any simulation-based method can at best provide a simplified view of real-world processes.
Consequently, we suggest augmenting the simulation model to incorporate measurement error in $y$.
Formally, we assume that there exists a measurement error (or \textit{noise}) model $g$ such that the true data generating process $p^* \in \{p(\cdot \mid \theta); \theta \in \Theta\}$ is given by
\begin{equation}\label{eq:misspecification}
    p(y \mid \theta) = \int_\mathcal{X} g(y \mid x) f(x \mid \theta) \, \d x \quad \text{and} \quad p(\theta \mid y) \propto p(y \mid \theta) p(\theta).  
\end{equation}
While the error model $g$ could in principle depend on $\theta$ as well, we do not consider this case in the remainder.
In practice of course, the \enquote{true} error distribution $g$ is also unknown, so that relatively coarse approximations will have to be used instead. 
\glsreset{gbi}
For this reason we advocate the use of \gls{gbi} \citep[e.g.][and Appendix \ref{appendix:gbi}]{bissiri2016general} within \gls{abc} to obtain robust posterior inference.

\subsection{A general view on likelihood-free inference}

\gls{gbi} extends the logic of Bayesian belief updating from a likelihood function to an arbitrary loss $\ell$.
Its main practical use lies in addressing robustness and misspecification \citep[see e.g.][]{AlphaDivPosterior, BetaDivPosterior, JackDivergences, RBOCPD}.
Replacing the measurement error distribution with a general loss $\ell$ and plugging it into the relevant \gls{abc} formulation yields
\begin{equation}
    p_\ell(\theta \mid y) \propto \int \exp{\{-w \cdot {\ell}(y; x)\}} f(x \mid \theta) p(\theta) \, \d x.
    \label{eq:general_posterior}
\end{equation}
The above generalizes standard \gls{abc}: instead of a probability kernel $K_h$, one may now use any loss function $\ell$ to assess the discrepancy between $x$ and $y$.
The weight $w > 0$ can be chosen to adjust the influence of a given loss on the posterior,  and we illustrate this in \autoref{fig:weights} using a toy example which is described below in Section \ref{sec:experiments}.

\begin{figure}[htbp]
\floatconts
  {fig:weights}
  {\caption{\Gls{abc} posteriors targeting \eqref{eq:general_posterior} using a \textit{soft-threshold} loss based on sufficient statistics $\eta(x)$, and a \textit{Wasserstein distance} loss on the full model output $x$.}}
  {%
    \subfigure[Soft-threshold \gls{abc}]{\label{fig:intro-soft-thresh}%
      \includegraphics[height=5.5cm]{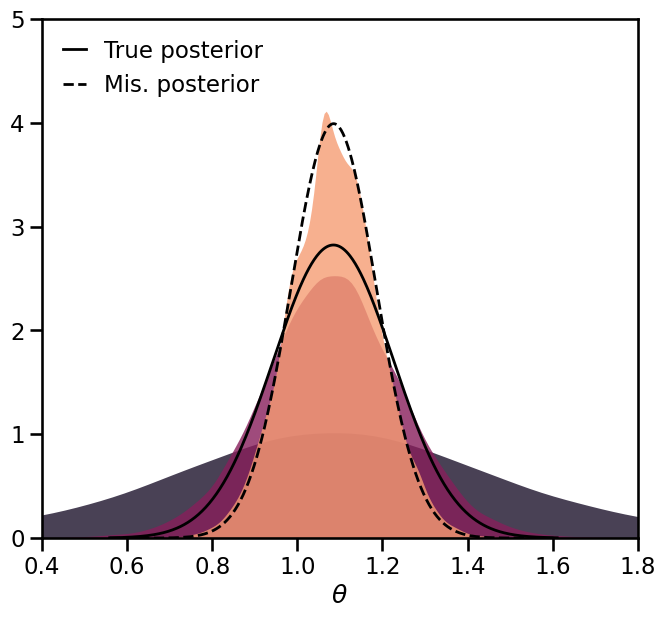}}%
    \qquad \qquad 
    \subfigure[Wasserstein \gls{abc} \hspace{8mm}]{\label{fig:intro-wass}%
      \includegraphics[height=5.5cm]{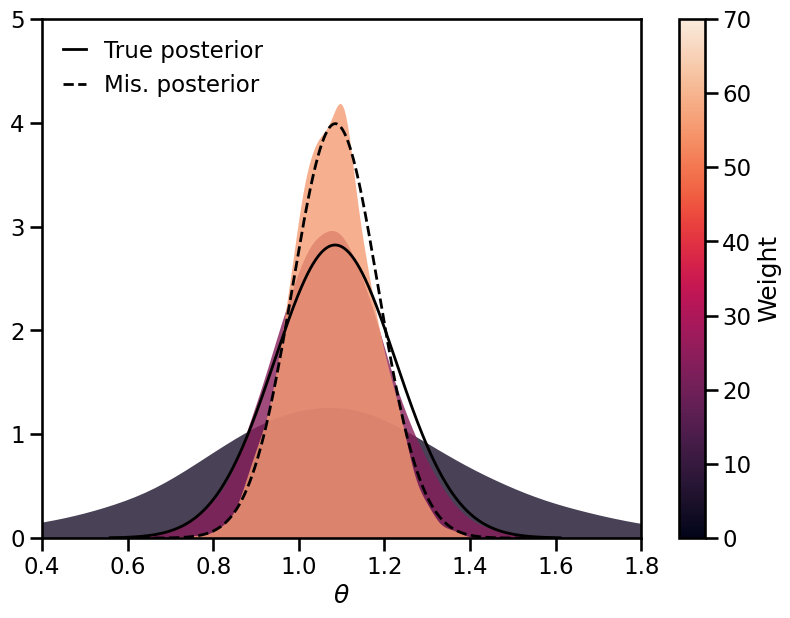}}
        \vspace*{-0.6cm}
  }
\end{figure}
\vspace*{-0.6cm}

\subsection{Links to approximate Bayesian computation}
\label{sec:linking_to_abc}
If the error distribution $g(y \mid x)$ of \eqref{eq:misspecification} is identical to the kernel $K_h(\|\eta(x)-\eta(y)\|)$, the corresponding \gls{abc} algorithm recovers the \textit{correct} posterior based on the likelihood $p(y\mid\theta)$ (rather than $f(y\mid\theta)$). 
While this argument for interpreting the usual \gls{abc} kernels as error distributions was previously made by \citet{wilkinson2013approximate}, it seems implausible that the correct measurement error of a complex simulator can be expressed in terms of a difference of low-dimensional summary statistics via $K_h$.  
Regardless, under this assumption one can substitute into \eqref{eq:general_posterior} the unit weight $w=1$ and the likelihood
\begin{equation}\label{eq:abc_kernel_to_loss}
    \ell(y; x) = - \log K_h(\lVert \eta(x)-\eta(y) \rVert)
\end{equation}
to recover not only the \gls{abc} posterior $p_{\gls{abc}}(\theta \mid y)$, but also the \textit{correct} posterior based on the likelihood  $p(y\mid\theta)$ of \eqref{eq:misspecification}. 
Doing so also demonstrates that the set of possible \gls{gbi} posteriors in \eqref{eq:general_posterior} contains all standard as well as a range of new and \textit{generalized} \gls{abc} posteriors, suggesting the traditional form \eqref{eq:ABC_generic} of \gls{abc} posteriors is an artificial restriction. 

Some popular \gls{abc} kernels gain new interpretations viewed through this lens.
For example, $K_h$ is often taken to be a Gaussian kernel, implying through \eqref{eq:abc_kernel_to_loss} the loss ${\ell}(y; x) \propto \frac{1}{2h^2}\lVert \eta(x) - \eta(y) \rVert ^2$.
This is a reasonable choice for $\ell$ whenever the (additive) observation noise on the summary statistics is at least approximately $\mathcal{N}(0,h^2)$ distributed.
On the other hand, if $\mathcal{N}(0,h^2)$ is a poor description of the error model, then \gls{gbi} robustification strategies could be deployed within \gls{abc} \citep[e.g.][]{RBOCPD, boustati2020generalized}. 
In contrast, the appeal of many popular \gls{abc} kernels is greatly reduced when viewed in this light.
For example, \textit{hard-threshold} kernels given by indicator functions such as $K_{\varepsilon}(x,y) = \mathbbm{1}\{ \lVert \eta(x) - \eta(y) \rVert < \varepsilon \}$ would correspond to a compactly supported and discontinuous noise model $g(y \mid x) = K_{\varepsilon}(x,y)$.
In virtually any other application, this choice of error model would be unconventional---yet it is commonplace in \gls{abc}.

We submit that \gls{abc} can elegantly address misspecification by leveraging \gls{gbi}:
Under the \gls{gbi} paradigm, we make more reasonable choices for $K_h$---choices that more accurately reflect the measurement error $g(y \mid x)$---that would be unnatural or even forbidden by the functional form of $K_h$ within standard \gls{abc}.
While some pre-existing work on \gls{abc} has taken small steps towards special cases of this generalization, these approaches do not consider the framework in its full generality
\citep[see e.g.][]{park2016k2, ridgway2019probably}.

\subsection{Sampling from ABC posteriors}

Given a choice of loss $\ell(y; x)$, \gls{gbi} requires us to compute expectations with respect to the posterior \eqref{eq:general_posterior}. 
The most common approaches are sampling algorithms based on the pseudo-marginal method (\citealp{beaumont2003estimation, andrieu2009pseudo}).
Such samplers follow a \gls{mcmc} approach: given a point $\{\theta, \ell(y; x)\}$, propose $\theta^\prime \sim \mathcal{N}(\cdot ; \theta, \lambda)$---a Gaussian centered at $\theta$ with variance $\lambda$---and $\ell(y; x^\prime)$ with $x^\prime \sim f( \cdot \mid \theta)$ and accept with probability
\begin{equation*}
    a(\theta, \theta') = \min\left\{1, \frac{ e^{-w \cdot {\ell}(y; x')}p(\theta^\prime)}{e^{-w \cdot {\ell}(y; x)}p(\theta)} \right\},
\end{equation*}
otherwise remain in the same state. An important result in \gls{mcmc} theory is that the respective Markov chain will indeed converge to the desired target distribution \eqref{eq:general_posterior}.
The resulting algorithm can be seen as a generalization of the \gls{abc}-\textsc{mcmc} scheme of \citet{marjoram2003markov}.

While we restrict our attention to \textsc{mcmc} methods, the suggested \gls{gbi} target \eqref{eq:general_posterior} is amenable to other popular \gls{abc} algorithms, e.g. importance sampling as proposed by \citet{park2016k2} or \textsc{smc}-\gls{abc} \citep{sisson2007sequential} using, for instance, the methods of %
\citet{fearnhead2008particle, bernrace}.

\section{Simulation Study}
\label{sec:experiments}

Throughout, we study the toy example of \citet{frazier2020}.
In this example, we assume that the true generative model comes from the family with likelihood $p(x \mid \theta) = \mathcal{N}(x; \theta, 2)$, that is, a Gaussian density with %
mean $\theta$ and variance $2$.
We generate 100 data points according to an unobserved $\theta^* = 1$, i.e. $p^*(y) = p(y \mid \theta^*)$, where our prior beliefs about this parameter are given by $p(\theta) = \mathcal{N}(\theta; 0, 25)$.
Instead of reflecting the correct likelihood model, our  simulator is misspecified and assumes that $y$ was generated from $f( \cdot \mid \theta) = \mathcal{N}(x; \theta, 1)$.
Note that this means the true error has a distribution given by $g(y \mid x)=\mathcal{N}(y; x, 1)$.

Our aim is to investigate the impact of the loss function on performance, as a judicious choice can not only help target the true posterior distribution, but may also substantially reduce the computational burden. 
This highlights the use of \gls{gbi} within \gls{abc} for improved robustness and computational efficiency as outlined in Section \ref{sec:linking_to_abc}.

Throughout, we focus on \gls{abc}-\gls{mcmc} algorithms \citep[see e.g.][]{fan2018abc} with soft-threshold losses, as we found this approach to perform very reliably. 
Where summary statistics are used, we employ the mean and variance $\eta = (\eta_1,\eta_2)$, which are \textit{sufficient} for the normal distribution.
The losses we consider take the form $\ell (y ; x) = \frac{1}{M}\sum_{i=1}^M \tilde{\ell}(y ; x_i)$ where $x_i \overset{iid}{\sim} f(\cdot \mid \theta)$  and $M$ denotes the number of simulator draws or \textit{particles}.
For $\tilde{\ell}$, we consider: the squared 2-norm applied to a difference of sufficient statistics, i.e. $\lVert \eta(x_i) - \eta(y) \rVert^2$, which we call \textsc{st}-\gls{abc}; the \gls{mmd} loss as used in \textsc{k}2-\gls{abc} \citep[][]{park2016k2}; and the Wasserstein distance, which we call \textsc{w}-\gls{abc} \citep[note that our approach differs from that of][]{Bernton_2019}.

\subsection{Solving the exact problem}

Formulating simulation-based inference as a latent variable model as we propose in \eqref{eq:misspecification} might suggest that the aim of \gls{abc} lies in an accurate modelling of the error distribution $g(y \mid x)$, i.e. finding $w\cdot \ell(y; x) \approx - \log g(y \mid x)$.
However, in practice considerations such as computational cost and robustness may well trump the benefits of modelling $g(y \mid x)$ exactly.
In order to test this hypothesis, we leverage the fact that the true error distribution in our toy model is known and employ a pseudo-marginal algorithm \citep{beaumont2003estimation, andrieu2009pseudo} to sample \emph{exactly} from the \textit{correct} posterior $p(\theta \mid y)$.

We tune this algorithm so as to maximize the \gls{ess} (using the \texttt{ess} function from the \textsf{ArviZ} library, \citealt{arviz_2019}) given a fixed budget of draws from the simulator.
In particular, we choose the number of draws per iteration \citep[this turns out to be more efficient than taking just a single sample, see e.g.][]{pitt2012some, doucet2015efficient, schmon2020large} so that the acceptance rate is around $25\%$ \citep{schmon2020large}.
As a comparison we take the \textsc{st}-\gls{abc} algorithm described above. 
The \gls{abc} algorithm uses an average of losses based on 15 samples, which improves the \gls{ess} for a given number of calls to the simulator (see \autoref{fig:robust_ess_experiment}), and is tuned so that it approximately targets the true posterior, as shown in \autoref{fig:abc_pmcmc} in Appendix \ref{appendix:links}. 
Further, we carry out a two sample Kolmogorov--Smirnov test on a subset of samples obtained from both methods, which demonstrates no significant difference between the distributions.
Given a budget of 1.5 million simulator calls, we compute the \gls{ess} of both the pseudo-marginal chain and the \gls{abc} chain. 
While the pseudo-marginal algorithm achieves an \gls{ess} of $\approx 2723$, \gls{abc}-\gls{mcmc} results in an \gls{ess} of more than $11080$---making it more than four times as efficient.
This has an important implication: while modelling the error distribution exactly may target the correct posterior, computational considerations may still favour alternative approaches.

\subsection{General loss functions}

The generalized posterior \eqref{eq:general_posterior} 
allows for consideration of a wide variety of losses with diverse properties.
We compare a selection of losses against the following criteria:
\begin{itemize}
    \item[i)] \emph{Computational cost.} Calls to most simulators of interest are expensive. Algorithms should make efficient use of simulations, perhaps through use of $M>1$ particles. 
    \label{itemize:criterion1}
    \\[-0.7cm]
    \item[ii)] \emph{Misspecification robustness.} Algorithms offering stable performance under high degrees of misspecification are to be preferred.%
    \label{itemize:criterion2}
    \\[-0.5cm]
\end{itemize}

\begin{figure}[htbp]
\floatconts
  {fig:robust_ess_experiment}
  {\caption{Two comparisons of \gls{abc} algorithms with a soft-threshold loss (\textsc{st}-\gls{abc}), an \gls{mmd} loss (\textsc{k}2-\gls{abc}), and a Wasserstein loss (\textsc{w}-\gls{abc}).}}
  {%
    \subfigure[The effective sample size as a function of $M$, the number of particles.]{\label{fig:ess_particles}%
      \includegraphics[height=5.98cm]{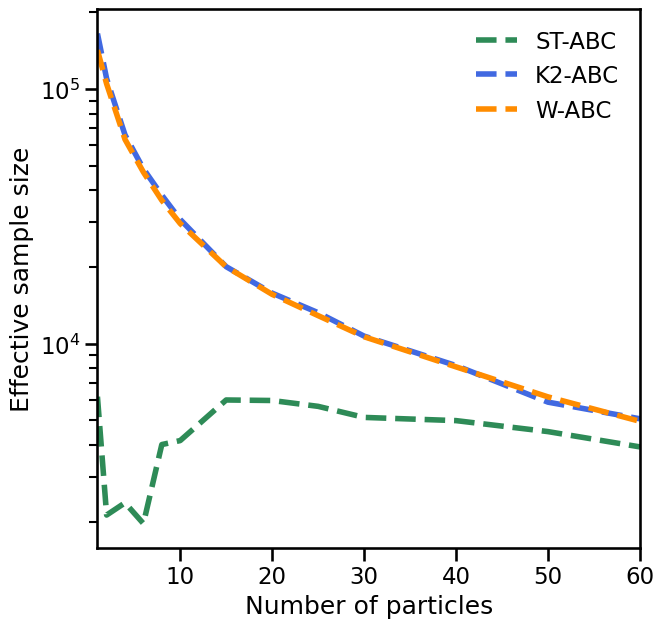}}
    \qquad \qquad
    \subfigure[The posterior mean as a function of the degree of misspecification of the model.]{\label{fig:robust}%
      \includegraphics[height=6cm]{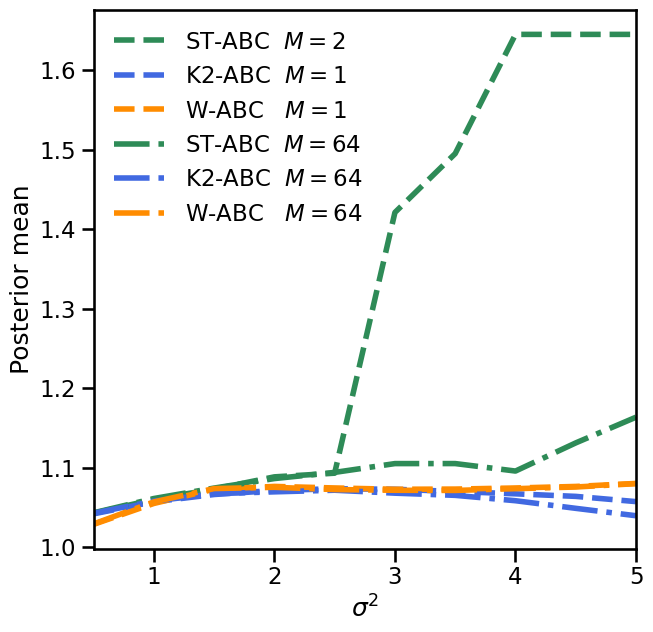}}%
    \vspace*{-0.5cm}
  }
\end{figure}

Criterion i) is addressed in \autoref{fig:ess_particles}. In this experiment, each algorithm run draws 3 million samples from the simulator, effectively fixing the computational cost. Those samples are spent on 3 million \gls{mcmc} iterations with $M=1$ particle, 1.5 million \gls{mcmc} iterations with $M=2$ particles, and so on. Finally, the efficiency is measured by the \gls{ess} of the \gls{mcmc} chain generated. For low numbers of particles, \textsc{k}2-\gls{abc} and \textsc{w}-\gls{abc} are vastly more efficient than the more conventional \textsc{st}-\gls{abc}. Both \textsc{k}2-\gls{abc} and \textsc{w}-\gls{abc} lose efficiency through any amount of loss-averaging. In contrast, the \gls{ess} of \textsc{st}-\gls{abc} is maximised at around 15-25 particles.
Turning to criterion ii), in \autoref{fig:robust} we explore the behaviour of the algorithms under misspecification. 
Here, we have assumed that the true data generating process is $\mathcal{N}(1, \sigma^2)$ and report the posterior mean of the \gls{abc}-\gls{mcmc} samples of each algorithm. Further details are provided in Appendix \ref{appendix:ex_details}. %
In agreement with \autoref{fig:ess_particles}, \textsc{st}-\gls{abc} produces posterior means of a far higher variance than the other losses, even using $M=2$ particles (see Appendix \ref{appendix:ex_details}).
Using 64 particles  improves performance for \textsc{st}-\gls{abc}---and yet, both \textsc{k}2-\gls{abc} and \textsc{w}-\gls{abc} still outperform it \textit{at every level of misspecification using only a single particle}.

\vspace*{-0.6cm}

\section{Discussion}
\glsreset{gbi}
\glsreset{abc}
We have introduced a unifying view for likelihood-free inference methods  by relating \gls{abc} to \gls{gbi}. 
Using the full expressivity of arbitrary loss functions allows us to re-interpret existing methods, but also suggests new promising directions for future research on \gls{abc}. 
We believe that this approach will broaden the scope of applicability for \gls{abc} algorithms, particularly in applications where simulators are known to be inexact representations of the real world data-generating mechanism for $y$.

\acks{
JK is funded by the EPSRC grant EP/L016710/1 as part of the Oxford-Warwick  Statistics Programme (OxWaSP). JK is additionally funded by the Facebook Fellowship Programme and the London Air Quality project at the Alan Turing Institute for Data Science and AI as part of the Lloyd's Register Foundation programme on Data Centric Engineering. This work was furthermore supported by The Alan Turing Institute for Data Science and AI under EPSRC grant EP/N510129/1 in collaboration with the Greater London Authority.
}

\bibliography{jmlr-sample}

\begin{thebibliography}{31}
\providecommand{\natexlab}[1]{#1}
\providecommand{\url}[1]{\texttt{#1}}
\expandafter\ifx\csname urlstyle\endcsname\relax
  \providecommand{\doi}[1]{doi: #1}\else
  \providecommand{\doi}{doi: \begingroup \urlstyle{rm}\Url}\fi

\bibitem[Andrieu et~al.(2009)Andrieu, Roberts, et~al.]{andrieu2009pseudo}
Christophe Andrieu, Gareth~O Roberts, et~al.
\newblock {The pseudo-marginal approach for efficient Monte Carlo
  computations}.
\newblock \emph{The Annals of Statistics}, 37\penalty0 (2):\penalty0 697--725,
  2009.

\bibitem[Andrieu et~al.(2010)Andrieu, Doucet, and
  Holenstein]{andrieu2010particle}
Christophe Andrieu, Arnaud Doucet, and Roman Holenstein.
\newblock Particle markov chain monte carlo methods.
\newblock \emph{Journal of the Royal Statistical Society: Series B (Statistical
  Methodology)}, 72\penalty0 (3):\penalty0 269--342, 2010.

\bibitem[Beaumont(2003)]{beaumont2003estimation}
Mark~A Beaumont.
\newblock {Estimation of population growth or decline in genetically monitored
  populations}.
\newblock \emph{Genetics}, 164\penalty0 (3):\penalty0 1139--1160, 2003.

\bibitem[Beaumont(2019)]{beaumont2019approximate}
Mark~A Beaumont.
\newblock {Approximate Bayesian computation}.
\newblock \emph{Annual review of statistics and its application}, 6:\penalty0
  379--403, 2019.

\bibitem[Beaumont et~al.(2002)Beaumont, Zhang, and
  Balding]{beaumont2002approximate}
Mark~A Beaumont, Wenyang Zhang, and David~J Balding.
\newblock {Approximate Bayesian computation in population genetics}.
\newblock \emph{Genetics}, 162\penalty0 (4):\penalty0 2025--2035, 2002.

\bibitem[Bernton et~al.(2019)Bernton, Jacob, Gerber, and Robert]{Bernton_2019}
Espen Bernton, Pierre~E. Jacob, Mathieu Gerber, and Christian~P. Robert.
\newblock {Approximate Bayesian computation with the Wasserstein distance}.
\newblock \emph{Journal of the Royal Statistical Society: Series B (Statistical
  Methodology)}, 81\penalty0 (2):\penalty0 235–269, Feb 2019.
\newblock ISSN 1369-7412.
\newblock \doi{10.1111/rssb.12312}.
\newblock URL \url{http://dx.doi.org/10.1111/rssb.12312}.

\bibitem[Bissiri et~al.(2016)Bissiri, Holmes, and Walker]{bissiri2016general}
Pier~Giovanni Bissiri, Chris~C Holmes, and Stephen~G Walker.
\newblock A general framework for updating belief distributions.
\newblock \emph{Journal of the Royal Statistical Society. Series B, Statistical
  methodology}, 78\penalty0 (5):\penalty0 1103, 2016.

\bibitem[Blum et~al.(2013)Blum, Nunes, Prangle, and Sisson]{Blum_2013}
M.~G.~B. Blum, M.~A. Nunes, D.~Prangle, and S.~A. Sisson.
\newblock {A Comparative Review of Dimension Reduction Methods in Approximate
  Bayesian Computation}.
\newblock \emph{Statistical Science}, 28\penalty0 (2):\penalty0 189–208, May
  2013.
\newblock ISSN 0883-4237.
\newblock \doi{10.1214/12-sts406}.
\newblock URL \url{http://dx.doi.org/10.1214/12-STS406}.

\bibitem[Boustati et~al.(2020)Boustati, Akyildiz, Damoulas, and
  Johansen]{boustati2020generalized}
Ayman Boustati, {\"O}mer~Deniz Akyildiz, Theodoros Damoulas, and Adam Johansen.
\newblock {Generalized Bayesian Filtering via Sequential Monte Carlo}.
\newblock \emph{arXiv preprint arXiv:2002.09998}, 2020.

\bibitem[Cranmer et~al.(2020)Cranmer, Brehmer, and Louppe]{cranmer2020frontier}
Kyle Cranmer, Johann Brehmer, and Gilles Louppe.
\newblock The frontier of simulation-based inference.
\newblock \emph{Proceedings of the National Academy of Sciences}, 2020.

\bibitem[Doucet et~al.(2015)Doucet, Pitt, Deligiannidis, and
  Kohn]{doucet2015efficient}
Arnaud Doucet, Michael~K Pitt, George Deligiannidis, and Robert Kohn.
\newblock {Efficient implementation of Markov chain Monte Carlo when using an
  unbiased likelihood estimator}.
\newblock \emph{Biometrika}, 102\penalty0 (2):\penalty0 295--313, 2015.

\bibitem[Fan and Sisson(2018)]{fan2018abc}
Y~Fan and SA~Sisson.
\newblock {ABC samplers}.
\newblock \emph{arXiv preprint arXiv:1802.09650}, 2018.

\bibitem[Fasiolo et~al.(2016)Fasiolo, Pya, and Wood]{fasiolo2016comparison}
Matteo Fasiolo, Natalya Pya, and Simon~N Wood.
\newblock A comparison of inferential methods for highly nonlinear state space
  models in ecology and epidemiology.
\newblock \emph{Statistical Science}, pages 96--118, 2016.

\bibitem[Fearnhead et~al.(2008)Fearnhead, Papaspiliopoulos, and
  Roberts]{fearnhead2008particle}
Paul Fearnhead, Omiros Papaspiliopoulos, and Gareth~O Roberts.
\newblock {Particle filters for partially observed diffusions}.
\newblock \emph{Journal of the Royal Statistical Society: Series B (Statistical
  Methodology)}, 70\penalty0 (4):\penalty0 755--777, 2008.

\bibitem[{Frazier, David T. and Robert, Christian P. and Rousseau,
  Judith}(2020)]{frazier2020}
{Frazier, David T. and Robert, Christian P. and Rousseau, Judith}.
\newblock Model misspecification in approximate bayesian computation:
  consequences and diagnostics.
\newblock \emph{Journal of the Royal Statistical Society: Series B (Statistical
  Methodology)}, 82\penalty0 (2):\penalty0 421--444, 2020.
\newblock \doi{https://doi.org/10.1111/rssb.12356}.
\newblock URL
  \url{https://rss.onlinelibrary.wiley.com/doi/abs/10.1111/rssb.12356}.

\bibitem[Ghosh and Basu(2016)]{BetaDivPosterior}
Abhik Ghosh and Ayanendranath Basu.
\newblock {Robust Bayes estimation using the density power divergence}.
\newblock \emph{Annals of the Institute of Statistical Mathematics},
  68\penalty0 (2):\penalty0 413--437, 2016.

\bibitem[Gr{\"u}nwald(2012)]{safeBayes}
Peter Gr{\"u}nwald.
\newblock {The safe Bayesian}.
\newblock In \emph{International Conference on Algorithmic Learning Theory},
  pages 169--183. Springer, 2012.

\bibitem[Guedj(2019)]{PACBayes}
Benjamin Guedj.
\newblock {A primer on PAC-Bayesian learning}.
\newblock \emph{arXiv preprint arXiv:1901.05353}, 2019.

\bibitem[Hooker and Vidyashankar(2014)]{AlphaDivPosterior}
Giles Hooker and Anand~N Vidyashankar.
\newblock Bayesian model robustness via disparities.
\newblock \emph{Test}, 23\penalty0 (3):\penalty0 556--584, 2014.

\bibitem[Jewson et~al.(2018)Jewson, Smith, and Holmes]{JackDivergences}
Jack Jewson, Jim~Q Smith, and Chris Holmes.
\newblock {Principles of Bayesian inference using general divergence criteria}.
\newblock \emph{Entropy}, 20\penalty0 (6):\penalty0 442, 2018.

\bibitem[Knoblauch et~al.(2018)Knoblauch, Jewson, and Damoulas]{RBOCPD}
Jeremias Knoblauch, Jack Jewson, and Theodoros Damoulas.
\newblock {Doubly Robust Bayesian Inference for Non-Stationary Streaming Data
  with $\beta $-Divergences}.
\newblock In \emph{Advances in Neural Information Processing Systems}, pages
  64--75, 2018.

\bibitem[Knoblauch et~al.(2019)Knoblauch, Jewson, and Damoulas]{GVI}
Jeremias Knoblauch, Jack Jewson, and Theodoros Damoulas.
\newblock Generalized variational inference: Three arguments for deriving new
  posteriors.
\newblock \emph{arXiv preprint arXiv:1904.02063}, 2019.

\bibitem[Kumar et~al.(2019)Kumar, Carroll, Hartikainen, and Martin]{arviz_2019}
Ravin Kumar, Colin Carroll, Ari Hartikainen, and Osvaldo~A. Martin.
\newblock {ArviZ} a unified library for exploratory analysis of {Bayesian}
  models in {Python}.
\newblock \emph{The Journal of Open Source Software}, 2019.
\newblock \doi{10.21105/joss.01143}.
\newblock URL \url{http://joss.theoj.org/papers/10.21105/joss.01143}.

\bibitem[Marjoram et~al.(2003)Marjoram, Molitor, Plagnol, and
  Tavar{\'e}]{marjoram2003markov}
Paul Marjoram, John Molitor, Vincent Plagnol, and Simon Tavar{\'e}.
\newblock {Markov chain Monte Carlo without likelihoods}.
\newblock \emph{Proceedings of the National Academy of Sciences}, 100\penalty0
  (26):\penalty0 15324--15328, 2003.

\bibitem[Park et~al.(2016)Park, Jitkrittum, and Sejdinovic]{park2016k2}
Mijung Park, Wittawat Jitkrittum, and Dino Sejdinovic.
\newblock {K2-ABC: Approximate Bayesian Computation with Kernel Embeddings}.
\newblock volume~51 of \emph{Proceedings of Machine Learning Research}, pages
  398--407. PMLR, 2016.

\bibitem[Pitt et~al.(2012)Pitt, dos Santos~Silva, Giordani, and
  Kohn]{pitt2012some}
Michael~K Pitt, Ralph dos Santos~Silva, Paolo Giordani, and Robert Kohn.
\newblock {On some properties of Markov chain Monte Carlo simulation methods
  based on the particle filter}.
\newblock \emph{Journal of Econometrics}, 171\penalty0 (2):\penalty0 134--151,
  2012.

\bibitem[Ridgway(2017)]{ridgway2019probably}
James Ridgway.
\newblock {Probably approximate Bayesian computation: nonasymptotic convergence
  of ABC under misspecification}.
\newblock \emph{arXiv preprint arXiv:1707.05987}, 2017.

\bibitem[Schmon et~al.(2020)Schmon, Deligiannidis, Doucet, and
  Pitt]{schmon2020large}
S~M Schmon, G~Deligiannidis, A~Doucet, and M~K Pitt.
\newblock {Large-sample asymptotics of the pseudo-marginal method}.
\newblock \emph{Biometrika}, 07 2020.
\newblock ISSN 0006-3444.
\newblock \doi{10.1093/biomet/asaa044}.
\newblock URL \url{https://doi.org/10.1093/biomet/asaa044}.

\bibitem[Schmon et~al.(2019)Schmon, Doucet, and Deligiannidis]{bernrace}
Sebastian~M. Schmon, Arnaud Doucet, and George Deligiannidis.
\newblock {Bernoulli Race Particle Filters}.
\newblock In Kamalika Chaudhuri and Masashi Sugiyama, editors,
  \emph{Proceedings of Machine Learning Research}, volume~89 of
  \emph{Proceedings of Machine Learning Research}, pages 2350--2358. PMLR,
  16--18 Apr 2019.
\newblock URL \url{http://proceedings.mlr.press/v89/schmon19a.html}.

\bibitem[Sisson et~al.(2007)Sisson, Fan, and Tanaka]{sisson2007sequential}
Scott~A Sisson, Yanan Fan, and Mark~M Tanaka.
\newblock Sequential monte carlo without likelihoods.
\newblock \emph{Proceedings of the National Academy of Sciences}, 104\penalty0
  (6):\penalty0 1760--1765, 2007.

\bibitem[Wilkinson(2013)]{wilkinson2013approximate}
Richard~David Wilkinson.
\newblock {Approximate Bayesian computation (ABC) gives exact results under the
  assumption of model error}.
\newblock \emph{Statistical applications in genetics and molecular biology},
  12\penalty0 (2):\penalty0 129--141, 2013.

\end{thebibliography}

\appendix

\clearpage

\section{Generalized Bayesian Inference} 
\label{appendix:gbi}
\glsresetall

\Gls{gbi} is a family of Bayes-like methods based around Gibbs posteriors.
These methods can be justified from multiple angles, including the PAC-Bayesian perspective \citep[see][for a recent survey]{PACBayes}, the so-called \textit{Safe Bayes} motivation of \citet{safeBayes}, a discrepancy-based perspective on Bayesian inference \citep[see][]{JackDivergences} as well as an optimization-centric view \citep[see][]{GVI}.
Perhaps most in line with the traditional Bayesian view is the conceptual justification of \citet{bissiri2016general}.
Under a restrictive definition of \textit{coherent belief updates}, the authors show that a Bayes-like update rule is a valid inferential device for general loss functions.
In particular, given a loss function $\ell: \Theta \times \mathcal{X} \to \mathbb{R}$ with parameter space $\Theta$ and a prior belief $p(\theta)$ over $\Theta$, one can show that a coherent belief update in the presence of an observation $y \in \mathcal{X}$ is given by the Bayes-like operation
\begin{equation*}
    p_\ell(\theta \mid y) \propto \exp(-\ell(y; \theta)) p(\theta).
\end{equation*}
Here, $p_\ell(\theta \mid y)$ is a \textit{coherent} belief whenever the normalization constant $\int \exp(-\ell(y; \theta)) p(\theta)\mathrm{d}\theta$ is finite.
This framework is appealing as it provides a new conceptual justification for belief distributions that have been studied elsewhere, particularly in the PAC-Bayesian literature.

\section{Further details to simulation study}
\label{appendix:ex_details}
In \autoref{fig:robust} the true data generating process is a $\mathcal{N}(1, \sigma^2)$, while the model is $\mathcal{N}(\theta, 1)$. Each experiment comprises drawing a new observation $y_{\sigma^2}=1+\sigma \xi$, where $\xi \sim \mathcal{N}(0, 1)$; the seed is held fixed so that $\xi$ is constant across observations. Then $3 \times 10^6$ \gls{mcmc} samples are generated from the respective sampler. The dashed lines show the performance for a single particle, with the exception of soft-threshold \gls{abc}, which exhibited too high variance in the \gls{ess} estimate with only a single particle and so two particles were used. Even with two particles, for reasonable levels of misspecification ($\sigma^2 > 2.5$) the \gls{ess} for \textsc{st}-\gls{abc} diminished extremely quickly. Thus, not only do the posterior means drift further from the true value as misspecification increases, but the algorithm itself ceases to function reliably. This shortcoming was not present in any other algorithm tested.

In advance of running the experiments in \autoref{fig:robust_ess_experiment}, each algorithm was tuned by setting its weight to approximately target the true posterior. In that way, every algorithm tested could have been used to perform the same inferential task.
Once the weights were set, to verify the algorithms were producing samples approximately from the same distribution, sub-samples of length 150 were randomly sampled from each posterior chain and compared using a two-sample Kolmogorov--Smirnov test. As the null hypothesis was not rejected at the 0.1 level for 5 different observations $y$, we found the posteriors satisfactorily close.
\begin{figure}[htbp]
\floatconts
  {fig:abc_pmcmc}
  {\caption{
  \Gls{abc} and a pseudo-marginal algorithm compared to the true posterior. The sampled posteriors are very close to the intended target distribution.
  }}
  {%
    \includegraphics[width=.6\textwidth]{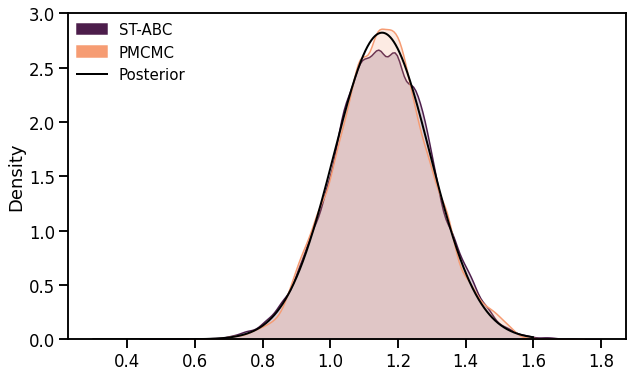}
    \vspace*{-0.6cm}
  }

\end{figure}

\section{Links to hidden Markov models} \label{appendix:links}

The view of a latent variable model with measurement error is one that is commonly used in the field of hidden Markov models.
Following our previous notation, models consist of a hidden Markov chain evolving according to
\begin{equation}
    x_i \mid x_{i-1} \sim f(\cdot \mid x_{i-1}, \theta), \quad y_{o, i} \mid x_i \sim g(\cdot \mid x_i, \theta).
\end{equation}
Denote the observed time series $y = (y^{(1)}, \ldots, y^{(n)})$, we can then recover expression \eqref{eq:general_posterior} by setting
\begin{equation*}
    p(\theta \mid y) = \int_\mathcal{X} g_\theta(y \mid x) f(x \mid \theta)\d x \, p(\theta), 
\end{equation*}
with $ f_1(x^{(1)}\mid x^{(0)}, \theta) =f_1(x^{(1)}\mid \theta)$ and 
\begin{align*}
f(x \mid \theta) = \prod_{i=1}^n f_i(x^{(i)}\mid x^{(i-1)}, \theta), \quad 
g_\theta(y \mid x) = \prod_{i=1}^n g_i(y^{(i)} \mid x^{(i)}, \theta).
\end{align*}
Particle Markov chain Monte Carlo algorithms \citep[][]{andrieu2010particle} are then designed to target posteriors like \eqref{eq:misspecification}. 
Accordingly, they are a possible solution for inference with \gls{gbi} posteriors whenever the simulator is formulated as a hidden Markov model with known error. 

It is noteworthy that for dynamic models \gls{pmcmc} and \gls{abc} are often competitors for solving the same task \citep[e.g.][]{fasiolo2016comparison}. However, it seems unreasonable to assume that the simulator is well-specified when employing one set of techniques (e.g. \gls{abc}) while assuming a measurement error model (e.g. \gls{pmcmc}).
In this sense, we suggest to decompose \gls{abc} analogously to hidden Markov models into signal and noise and to consider \gls{abc} with appropriately chosen loss as a \emph{robust} inference method. 
This is also in line with the results of \citet{fasiolo2016comparison} who find that 
\gls{pmcmc} performs better when the model and noise are well-specified, but may underperform under model misspecification. 
Perhaps this could be alleviated using  robust loss functions in the vein of \eqref{eq:general_posterior} on their individual noise contributions 
as proposed \citet{boustati2020generalized}.

\end{document}